# WERNER HEISENBERG[*]

(1901-1976)

Ivan Todorov
Institut für Theoretische Physik, Universität Göttingen, Friedrich-Hund-Platz 1
D-37077 Göttingen, Germany. E-mail: itodorov@theorie.physik.uni-goe.de
and Institute for Nuclear Research and Nuclear Energy, Tsarigradsko Chaussee 72,
BG-1784 Sofia, Bulgaria. E-mail: todorov@inrne.bas.bg

**Abstract.** A brief review of Heisenberg's life and work: participating in the youth movement in the aftermath of World War I, creating quantum mechanics, conflict with "deutsche Physik", involvement in "Hitler's Uranium Project", last illusions. Problems and dilemmas for scientists under a dictatorship – East and West.

Years ago, at the Symposium on Contemporary Physics in Trieste, in June 1968, a colleague from Israel left the audience when Heisenberg was to start his lecture – to demonstrate his contempt of a "Nazi collaborator". Samuel Goudsmit (1902-1978), the scientific head of the Alsos[1] mission, has also manifested strong negative feelings[2] in his 1947 book [G]. The passions are still alive: the insightful 1998 play [F], in which different versions of Heisenberg's 1941 meeting with Bohr in Germany occupied Copenhagen are rehearsed, provoked a heated discussion on the pages of Physics Today [PT] among historians and some over 90 years old witnesses of those days[3].

The moral questions about Heisenberg's behaviour in the Nazi period lead us to the more general topic of the scientist under a dictatorship, a topic which concerns not only Germans. This motivates the focus in the present article. Of the many achievements of Heisenberg starting with his doctorate (with Sommerfeld) on turbulence, going through his seminal contributions to atomic and nuclear physics, to the foundations of quantum field theory and the theory of ferromagnetism, I shall only briefly touch upon his 1925 paper on the matrix formulation of quantum mechanics and mention his late attempt to construct a unified quantum field theory involving a fundamental length. I will lay instead more emphasis on the personality of Heisenberg, on his early experiences, and especially on his encounter with political issues after Hitler came to power in 1933.

The author thanks Professor Klaus Gottstein, a long-term colleague of Heisenberg's in Munich, for a critical reading of an earlier draft of the paper and for valuable information.

---

[*] Extended version of a lecture presented at the Conference, dedicated to the 50th anniversary of the re-establishment of the Alexander von Humboldt Foundation, held in Sofia, Bulgaria, October 2003.
[1] Greek for grove, as in General Leslie Groves, the military head of the Manhattan Project; a scientific intelligence mission that followed the allied troops in the wake of their invasion to Europe.
[2] Later Goudsmit expressed regret for what he wrote, both publicly and to Heisenberg's son Jochen (see J. Heisenberg's talk at the Symposium "The Copenhagen Interpretation: Science and History on Stage", Washington, D.C. 2002, available at http://web.gc.cuny.edu/ashp/nml/artsci/heisenberg.htm).
[3] The disputes were not resolved by the release of the draft of an unsent letter of Bohr (along with other documents) by the Niels Bohr Archive (the polemic is reflected in [Got] where a convincing defence of Heisenberg's point of view on the matter is presented ) . For continuing discussions (on this and related topics) – see [NYR] and [Got-B], as well as the review [Oe].

1. **Pathfinders' leader**

Werner Heisenberg is the ambitious son of ambitious parents. His maternal grandfather, headmaster of the prestigious Maximilian Gymnasium in Munich, which the boy attends, stimulates his desire to be the first at all subjects. Having no problem with languages and science, Werner runs every evening around the local park, a stopwatch in hand, achieving eventually top grades in sports too (see [EH], Chapter 6). As he displays an early interest in number theory, his father, a professor in Byzantine studies, brings him a work in Latin by the famous number theorist L. Kronecker (1823-1891).

During the war he takes part in a Military Preparedness Association in school. In the spring of 1919, after communist victories in Hungary and Austria, when the German followers of Lenin seize control in Munich proclaiming a Bolshevik Soviet Republic, Werner's unit sides with the forces of Berlin's socialist war minister that crash the "military dictatorship of the proletariat"[4]. After the war Heisenberg is an active participant in the Bavarian youth movement: he becomes the leader (*Führer*) of a group of Pathfinders (*Pfadfinder,* the German version of Boy Scouts founded before the war). Half a century later he will start his recollections [H] with these experiences. In the chaos of the defeated postwar country educated youths of good family "drew together in larger and smaller groups in an attempt to blaze new paths, or at least to discover a new star to steer by" [H]. They turn, in fact, to the romantic ideals of the past, to harmony with Nature, to music and philosophy, far from the dirty world of politics and big money. Some authors[5] argue that such an elite apolitical stand has left a vacuum filled later with extremism and demagogy. I am not sure that anybody would have gained much had these young romantics indulged in politics instead[6]. In any case, long excursions with friends of his youth group serve Heisenberg to relax after periods of intense work and he keeps his ties with his "pathfinders" friends for many years.

In the fall of 1920 Heisenberg enters the Munich University. His early achievements match the expectations of his parents and young followers. Even an apparent failure at the beginning turns beneficial to him. Werner opts for studying pure mathematics but old Ferdinand von Lindemann (1852-1939), famous for the proof of the transcendence of π, sends him away (each time the eighteen-year-old boy tries to say something the poodle of the half-deaf professor barks at him... – cf. [C] p.99). So, Heisenberg becomes a student of Arnold Sommerfeld (1868-1951) who comes to teach more than one Nobel Prize winner. Werner is included from the outset in the research program on atomic structure of his teacher's. When in the summer of 1922 he leaves to the United States for a year, Sommerfeld sends Heisenberg to Göttingen, the Mecca of mathematical physics in Germany at the time. There, the second year physics student also meets Niels Bohr (1885-1962) in the wake of his arrival from Copenhagen. Bohr is visiting – in June 1922 – the small University town for a lecture series[7]. Thus, at 20, the youngster is introduced to the three world centres of theoretical atomic physics of the 1920's: Munich, Göttingen, and Copenhagen.

---

[4] Heisenberg's biographer [C] (Chapter 4) is not happy that upper middle class Munich intellectuals side with right wing forces (which include Hitler) in suppressing the Soviet Republic albeit he does not spare the fact that the regime of red terror relied solely on a well paid army and provoked a collapse of the economy.

[5] See references to G.L. Mosse, *Nationalism and Sexuality*, 1985, and to H. Pross, *Jugend-Eros-Politik*, 1964 in [C].

[6] My own father, Todor Borov (1901-1993), did engage in left wing political activity as a Bulgarian student in Berlin in the 1920's and had the presence of mind to express, as an old man, his regrets for not having pursued more vigorously his doctoral study instead.

## 2. Creating quantum mechanics

The moment of truth comes to Werner Heisenberg in June 1925. After a year in Copenhagen working together (and struggling) with Hendrik Kramers (1894-1952) and Wolfgang Pauli (1900-1958), another student of Sommerfeld's, in the young entourage of Bohr, he spends a month of inconclusive work (while lecturing) in Göttingen. Seeking relief from an attack of hay fever he ends up alone on the tiny island of Helgoland in the North Sea, off the German coast. There, away from busy theorists (and any other people for that matter), he has the decisive idea that opened the way to a consistent theory of the atom. He is excited but not quite sure of himself. While mailing a copy of his manuscript to Pauli (on July 9) and handing his only other one to Max Born (1882-1970), his Göttingen mentor, he writes to his father: "My own work is not going at the moment especially well...". He does not mention his latest results while lecturing in Cambridge, later in July. Born is, by contrast, fascinated by the article of his young assistant and forwards it to *Zeitschrift für Physik* while Werner is in England.

Let me try to describe in a few words the background of this discovery.

The quantitative study of atoms was made possible by the 19th century development of *spectral analysis*: it was recognized by 1860 (when a joint paper of Kirchhoff and Bunsen was published) that emission (and absorption) spectra of elementary substances characterize them completely[8]. A starting point for the subsequent study of atomic structure was Balmer's[9] 1885 formula for the frequency spectrum of the simplest of atoms, the hydrogen:

$$\nu_{mn} = \nu_m - \nu_n, \text{ with } \nu_n = R/n^2 \text{ (n>m).} \qquad (1)$$

The important feature of this relation (later extended to the spectra of arbitrary atoms by Rydberg, 1889, and Ritz, 1908) is that observed frequencies, labeled by *two* integers, can be presented as differences of two frequencies[10], each labeled by a single one. Max Planck (1858-1947) in his epoch making study of the *universal radiation law* for hot bodies (called the *black body radiation*) was led to postulate that the energy of light quanta is proportional to its frequency $\nu$ (with proportionality coefficient equal to the *Planck's constant* h):

$$E = h\nu. \qquad (2)$$

All this led Niels Bohr (in 1913) to the idea that an atom has a discrete set of *stationary states* whose energy is labeled by an integer and only emits light during the transition from a higher to a lower energy state, with frequency proportional to the energy difference. Moreover, Bohr computed the Rydberg constant in terms of the electron charge and mass (and Planck's constant). A major step forward [11] as it was, Bohr's model was hard to believe: the postulate of discrete stationary states ("electron orbits") was contradicting the laws of classical physics (as its author pointed out himself, a classical electron moving around a positive nucleus would loose energy by radiating and will eventually fall on the positive charge). On the other hand, it fitted beautifully the spectral data. The situation resembled the story of the horseshoe on the front door of Bohr's country house in Tisvilde: when asked whether he really had faith in such signs, Bohr used to say: "Of course not, but they say that it helps even non-believers." He is recorded to have buttonholed people (in Manchester in 1914) with the question "*Do you believe it?*" The forty-five-year-old Sommerfeld, Heisenberg's teacher, did (most other believers were younger): he welcomed enthusiastically the new development and set himself elaborating the fine structure of atomic spectra.

---

[7] After one of Bohr's lectures at which Werner makes a critical remark, Bohr invites the boy for a walk in the outskirts; then and there he invites Heisenberg to visit his Institute in Copenhagen. ("*By the way, what has happened to the life of scientists? Where have all those walks gone?*" exclaims G. Holton in [PT] on a similar occasion.)

[8] For a concise and illuminating survey of this development see Pais' book [P], Chapter 9.

[9] Johann Jakob Balmer, born in 1825 in Lausanne, studied mathematics in Germany, received a Ph.D. in Basel (a thesis "On cycloids") in 1849, and remained there as a teacher in a girls' school until his death in 1898. His *Habilitationsschrift* (of 1865) "*The prophet Ezekiel's vision of the temple, clearly portrayed and architecturally explained*" was devoted to biblical geometry (cf. [P], p.172). Balmer determined "Rydberg's" constant R in (1) with a surprising for an amateur accuracy of 1/10000.

[10] It follows that – in contrast to the classical physics (where frequencies form an abelian group) – not any two frequencies can be added (to obtain another observable frequency). This is only possible if the intermediate indices coincide; then $\nu_{mi} + \nu_{in} = \nu_{mn}$. The significance of this remark and its relation to the non-commutative geometry of the quantum mechanical phase space is underscored by A. Connes [Co].

[11] To appreciate Bohr's intuition one should look at prior attempts to interpret (the abundant – and seemingly confusing – data on) atomic spectra - see the vivid account in [P] pp. 197-199.

Heisenberg begins the abstract of his breakthrough Helgoland paper "*On a quantum theoretic reinterpretation of kinematical and mechanical relations*"[12] with the words: "*... it will be attempted to secure foundations for a quantum theoretical mechanics, based exclusively on relations between quantities which are, in principle, observable.*" The emphasis on *observable* (that was taken at face value by followers and commentators) was meant to exclude the motion of an electron on a stationary orbit whose understanding was causing problems. The really important point was the new multiplication rule proposed in the paper. The logic went roughly as follows. Atomic spectroscopy gives information about the transition between *two stationary states*. All physical quantities, not just the frequencies, should be labeled by *a pair of numbers*, say (m, n), corresponding to the two energy levels: $x_{mn}$, for the coordinate x; $p_{mn}$, for the momentum p. The product of two such arrays is given by a sum:

$$(xy)_{mn} = \Sigma_s x_{ms} y_{sn}.$$

Heisenberg will later learn from Born that this is the multiplication law for *matrices*, known to mathematicians. He recognizes the important fact that his product is not necessarily commutative, in general, $(xy)_{mn}$ differs from $(yx)_{mn}$, - and he is troubled by this observation. The first step in formulating the new mechanics is the most difficult one. The young author is both excited and hesitant. It is only in the subsequent work of Born with Werner's coeval, Pascual Jordan (1902-1980) and of a British youngster, Paul Dirac (1902-1984), in the fall of 1925, that the basic commutation relation between position q and momentum p is given,

$$[q,p] = qp - pq = i\hbar, \qquad (3)$$

and it is recognized that the Hamiltonian equations of motion remain the same as in classical mechanics (if one just substitutes Poisson brackets by commutators): only the algebra of physical quantities is changed (into a non-commutative one). This implies a change in the physical meaning of notions like position and momentum. It is Born who provided – in June 1926 - the definitive probabilistic interpretation of quantum mechanics. Next comes Heisenberg's most famous (albeit not his greatest) contribution: stimulated by discussions with Bohr and Pauli he publishes (alone) in March 1927 his notorious uncertainty relations.

It would be appropriate to end this summary with the words Dirac chose when introducing Heisenberg at the 1968 Trieste Symposium ([FLP] p.32): *Werner Heisenberg and I were young research students at the same time, about the same age, working on the same problem. Heisenberg succeeded where I failed. There was a large mass of spectroscopic data accumulated at the time and Heisneberg found the proper way of handling it. In doing so he started a golden age in theoretical physics, and for a few years after that it was easy for any second rate student to do first class work.*

### 3. **Professor in Germany under Hitler**

At the end of January 1933 Adolf Hitler, the chairman of the *National Socialist German Workers' Party* (NSDAP) that won most votes in the elections, is appointed Chancellor (by the president, Field Marshal von Hindenburg). (The humiliating Versailles' treaty and the economic crises led to a polarization of political forces in the country. E.L. Feinberg ([Fe], pp. 253-261) argues rather convincingly that the 14-fold increase of the Nazi vs. communist vote between 1928 and 1932 is provoked in part by Stalin's collectivization which affects the German settlers in Russia since the time of Catherine the Great – along with Russians and Ukrainians.) Hitler's words "*Faith and trust shall not be disappointed!*", pronounced at his first cabinet meeting, are welcomed by most Germans. A month later the Reichstag, the German parliament, is set on fire, allegedly, by the Dutch communist van der Lubbe, serving the Nazis as pretext for suppressing the political opposition.[13] German professors, people of unquestioned integrity like Max Planck, President of the Kaiser Wilhelm Society and secretary of the Prussian Academy, are trying to separate science from political struggles and the interference of ideology. When Einstein makes a public statement in the US (on March 11, 1933) that he would not live in Germany while people there are persecuted for their views, von Laue (1879-1960)[14], who is trying at the time, together with Planck, to improve the situation through quiet diplomacy, warns him: "Here they are making nearly the entirety of German academics responsible when you do something political."

---

[12] W. Heisenberg, *Über quantentheoretische Umdeutung kinematischer und mechanischer Beziehung*, Zeitschr. f. Phys. **33** (1925) 879-893, [WH] **AI**, pp. 382-396**;** for an English translation and historical background – see [W], pp. 261-276.

April 1933 starts with organized manifestations of anti-Semitism ("days of boycotting Jewish stores" in which storm troopers in brown shirts harass shop owners and their customers, provoke Jewish professionals and teachers) and ends with the enactment of laws forcing Jewish civil servants and University professors to retirement. The Propaganda minister Joseph Göbbels declares that the boycott and harassment of Jews will go on "until such anti-Nazi propaganda" (as Einstein's) "ceases".

The new laws strongly affect German Universities. Respectable physicists like Planck, von Laue and Heisenberg are doing their best to preserve for Germany scientists of the stature of Max Born, the Nobel laureates Fritz Haber (1868-1934) and James Franck (1882-1964), the Göttingen mathematician (Hilbert's associate) Richard Courant (1888-1972). The seventy-five-year old Planck uses his audience with Hitler to try to convince him that Jewish scientists could be good Germans mentioning, in particular, the case of the physical chemist Haber who helped Germany during World War I to introduce gas warfare. But to no avail (at a certain point the Chancellor flies into such a rage that Planck can do nothing but leave). The policy leading to the exodus of Jewish scientists is not altered. Born, barred from the classroom and taking refuge at his summer home in northern Italy, postpones for some time his final decision to leave Germany but, after his older children, left behind in Göttingen, experience further indignities, he accepts a three year appointment in Cambridge and resigns: "*One cannot serve a state that treats him as a second class citizen - and his children[15], even worse,*" he tells Sommerfeld. Hermann Weyl (1885-1955), whose wife is Jewish, and the Austrian Erwin Schrödinger (1887-1961), although not personally threatened, also leave Göttingen and Berlin, respectively, citing health and working conditions. Failing to keep their colleagues in Germany, Planck, von Laue, and Heisenberg are trying to at least to find reasonable replacements for the resulting vacancies. Here they meet a fierce opposition from their pro-Nazi experimental colleagues, in particular, the Nobel laureates Philipp Lenard (1862-1947) and Johannes Stark (1874-1957), who use the slogan of "German (or Aryan) physics" in order to get rid of the newly flourishing, with the birth of relativity and quantum mechanics, theoretical physics and to increase their personal influence. In order to be able to counter such a pressure, decent people had to keep their position and hence to face day after day the problem where to draw the line of compromise with the regime. Planck, at the age of seventy-six, has been recorded (by a foreigner present in the audience) to have repeatedly hesitated before raising his hand in the ritual "Heil Hitler" salute at the beginning of his speech at a meeting of the Kaiser Wilhelm Society[16]. Von Laue, who did not lecture, is said to have managed to avoid the salute by keeping his hands busy carrying something whenever going out... Heisenberg has to cope with the mechanical hand raising at the beginning of each lecture, but he never joins the NSDAP or any of its affiliated organizations (while 56% of the German nuclear physicists have been party members and 72% have participated in some Nazi organization – see [HH] and [L]; for instance, his younger colleague and life-long friend, von Weizsäcker, has been a member of the University Teachers League). Moreover, he refuses to participate in a highly publicized November 1933 rally of the National Socialist Teachers League, held in his University town, Leipzig, in support of Germany's withdrawal from the League of Nations (while, e.g., the world famous existentialist philosopher M. Heidegger (1889-1976) is prominently present). A vindictive Stark, the rally organizer, incites Nazi students to obstruct Heisenberg's lecture, the day after the announcement of his Nobel Prize (the same month). Happily, a member of his youth group manages to thwart the action ([C] pp.323-324). After Hindenburg's death in early August 1934 a plebiscite is scheduled to approve the fusion of the offices of chancellor and president in the hands of Hitler. Stark invites his fellow Nobel laureates to join in a public declaration in support of Hitler. Among physicists, Heisenberg, von Laue, Planck, and Walter Nernst (1864-1941) refuse, referring again to the argument that science and politics should not mix.

---

[13] The victims of Nazi repressions for the 6 years before the war count nearly 10 000 people. Alas, this is not a record for the past 20th century: the number of killed in tiny Bulgaria during the first 10 years of communist dictatorship after September 9, 1944, is about 3 times bigger (while the victims of Lenin's terror – according to the official Soviet census of 1924 – exceed 18 millions, not including those killed during the civil war!).

[14] Max von Laue, 1914 Nobel Prize winner, is known for his dignified and courageous behaviour in Hitler's time.

[15] Born's son still cannot forgive Heisenberg for having, allegedly, invited his father to come back to Germany, during a visit to Cambridge in 1934, without offering protection to his family – see Gustav V.R. Born, *The Born Family in Göttingen and Beyond*, Göttingen Institut für Wissenschaftsgeschichte, 2002, p.51.

  Many Western observers, including D. Cassidy, the author of the best available, well documented "Life of Heisenberg" [C] (used here extensively), assert that "even the most upright among German scientists have been thoroughly unprepared" to meet the moral and political problems posed by the new regime ([C] p.312). Cassidy argues that "a mobilization of mass opposition could have been achieved in early 1933 had more non-Jewish academics followed the examples of Einstein, Franck, and Haber: that is, had there been a simultaneous resignation of professors in moral indignation of the dismissal of their colleagues and the treatment of Jews in general" ([C] p.306). I am not convinced that this judgment is correct. Cassidy, himself, cites the story of Leipzig economics professor G. Kessler who "was dismissed from his post in April 1933" and later "arrested by the Gestapo" and forced to "emigrate soon thereafter to the United States" as a result of his "public lectures against National Socialist doctrine" ([C] p. 305). When Pyotr Lebedev (1866-1912) - the patron of today's Academy Physical Institute in Moscow - together with 130 Moscow University professors resigns in protest against the expulsion of revolutionary students in 1911, the tsarist government is unmoved. The only affected are the students, interested in science, and Lebedev himself who dies a year later at the age of 46 (see [Fe]). By contrast, Pyotr Kapitsa (1894-1984) does not resign when Lev Landau (1908-1968) is arrested at his Institute for Physical Problems in 1937, but uses his position as a director to convince the authorities to free Landau a year later. An example outside the life of scientists is provided by the saving of Bulgarian Jews during World War II: it has been achieved through the effort of a number of Bulgarians channeled by the vice chairman of the National Assembly, D. Peshev, who has been careful to use the support of members of the government party only, helped by the Bulgarian Orthodox Church and eventually by Tsar Boris III - thus working within the authoritarian system, not in an open protest against it (see [T]).

---

[16] See [Fe] p. 237; a habit, hardly worse than the inescapable "applauding going into ovation" at every public mentioning of the name of Stalin (as well as that of the local leader) in the "countries of the real socialism".

4. **The Black Corps**

Just as the trial of Galileo has been initiated by fellow scientists and philosophers who rely on the power of the Church to get revenge upon a superior mind (see, e.g., [D]) the attack on modern physics in Nazi Germany is triggered by overjealous experimental physicists, Lenard and Stark, who cannot cope anymore with the new theoretical developments in their own field and try to use the official racist ideology to dismiss them as alien to German spirit (see [Be] and [HH]). Their attacks on the curriculum in theoretical physics (for which they rely on a Nazi physics student) and on Heisenberg personally (by Stark) are becoming vicious. The very content of what is being taught and hence the future of science in Germany is threatened. Heisenberg feels obliged to take some of his time, happily devoted to his beloved physics, in order to defend it publicly (in the ugly "world out there" which he prefers to avoid – see [C] p.331). He gives a plenary address *"Transformations in the foundations of natural science in recent time"* at a meeting of the prestigious Society of German Scientists and Physicians in September 1934. Relativity and quantum theory, he says "have been forced upon research by Nature in the attempt to carry the program of classical physics consistently to its end." Philipp Lenard's investigations on the photoelectric effect, he adds, were adequately interpreted only in Albert Einstein's light quantum theory ([C] p.337). (He allows himself to cite Einstein's name at the price of combining it in one sentence with the name of the Nazi physicist.) In early 1936, faced with continuing attacks, he manages to publish – with the help of highly placed sympathizers - a carefully worded defense of modern theoretical physics in the party newspaper *Völkischer Beobachter* (now without citing names). Further research in relativity and quantum theory, he writes, "from which perhaps the strongest influence on the structure of the entire spiritual life will arise, is one of the most important tasks for German youth." A *New York Times* editorial praises him for his courage ([C] p.352). Stark's response in the same issue of *Völkischer Beobachter*, backed by the party ideologue A. Rosenberg, reiterates his demand: "The type of physics Heisenberg defends ought no longer, as it has until now, exert a decisive influence on filling physics teaching chairs." The fight moves to the Reich's education ministry. Heisenberg, backed by experimentalists M. Wien (1866-1938) and H. Geiger (1882-1945), has the upper hand – for the moment. He does not fight the new power (which in the eyes of most Germans is taking the country out of the deep crises): he defends his profession from demagogues-careerists.

Heisenberg reduces his non-professional contacts during this period to a narrow circle of music lovers in Leipzig. After the departure to Berlin of his younger friend and assistant Carl-Friedrich von Weizsäcker (and the break of his romance with Carl-Friedrich's sister, Adelheid) in 1936, he feels particularly lonely. In November that year he writes to his mother: "The single life is bearable to me only through my work in science, but in the long term it would be very bad if I had to make do without a very young person next to me." ([C] p.367) In late January 1937, the 35 years old

Heisenberg gets acquainted with the twenty-two-year old Elisabeth Schumacher (whose brother Fritz is an economist who later writes the now famous *"Small is Beautiful"*). It happens at one of his musical evenings in a publisher's home. Within two weeks they are engaged and marry in Berlin in less than 3 months. (Nine months later Elisabeth gives birth to fraternal twins, Anna Maria[17] and Wolfgang (named after Pauli who congratulates the happy father for his "pair creation"); she will bear Heisenberg five more children, four of them before 1944.)

In mid-July of the same year (1937) the couple arrives in Munich where Werner's mother lives and everything is prepared for getting Sommerfeld's chair – according to the long expressed wish of Heisenberg's retiring teacher. Bad news is awaiting him there, however: the SS weekly *Das Schwarze Korps* (The Black Corps – alluding to the black uniform of the SS troops) of July 15 contains a full-page attack on science and on Heisenberg entitled *"'Weisse Juden' in der Wissenschaft"* ("White Jews" in Science), signed by Stark. "It is not only the racial Jew in himself who is a threat to us, but rather the spirit he spreads.", the article reads, "And if the carrier of this spirit is not Jew but a German, then he should be considered doubly worthy of being combated as the racial Jew, who cannot hide the origin of his spirit. Common slang has coined a phrase for such bacteria carriers, the 'white Jew'." After vilifying "the Jews Einstein, Haber, and their like-minded comrades, Sommerfeld and Planck," and accusing them again of manipulating physics appointments to exclude "Germans" the article is mounting a full scale assault on "that white Jew ... representative of the Einstein 'spirit' in the new Germany ... Heisenberg" (see [C] pp.379-381)[18]. He is presented as a covert state enemy whose crimes (enumerated meticulously) include "smuggling" his article defending teaching of relativity in the party newspaper, refusing to join his fellow Nobel Prize winners in the 1934 declaration of support of Hitler's presidency, dismissing a German assistant in his institute in favour of the Jewish physicists Bloch and Beck... More ominously, a large print subtitle is calling Heisenberg the "Ossietzky" of physics. (Carl von Ossietzky (1888-1938) is a German journalist and pacifist, who refuses to flee Germany when Hitler comes to power. He is arrested in the end of February 1933 and sent to Papenburg concentration camp, transferred in May 1936 to a prison hospital. He is awarded in November 1936 the Nobel Peace Prize for 1935; this prompts Hitler to issue a decree forbidding Germans to accept any Nobel Prize.) "Representatives of Judaism in German spiritual life must all be eliminated as the Jews themselves", the article says. Stark and his people have struck revenge. They have found a serious connection: Hitler's elite servicemen.

To the credit of German scientists, Heisenberg's colleagues and responsible academics do not bow to the threat but courageously defend him – and their science. Munich rector Kölbl transfers a letter of protest by Sommerfeld to the Bavarian Culture Ministry adding his own remark: "It is outrageous that an active professor, a civil servant of the National Socialist state, should be attacked in this way." Leipzig's colleague Friedrich Hund writes letters in support to both the rector and to the Reich's Education Minister. In the months immediately after the Black Corps' article the Göttingen Academy of Sciences elects Heisenberg corresponding member and the Saxon Academy elects him deputy secretary of the math and physics class ([C] 382). Had Heisenberg sought at this moment emigration the SS paper's accusations would have worked as an asset for him. But he would have betrayed his supporters if he used the possibility without an attempt to stop the destruction of German physics. So he starts a fight to clear himself from the accusations that appear (at least for the outside world) to honour him. He writes a letter to SS Reichsführer, H. Himmler, asking him to decide: either to approve of Stark's attack – in which case Heisenberg would resign, or disapprove of it, in which case he demanded "restitution of his honour" and protection from further attacks. His mother knows Himmler's mother (who also lives in Munich) and transmits her son's letter through her. A painful SS investigation is started that lasts more than eight months. Heisenberg even has to go for an official interrogation to the notorious basement chambers of the SS headquarters at Prinz-Albert-Strasse 8 in Berlin. In spite of the favourable outcome, this investigation leaves a heavy imprint on him to the end of his life. Long after the regime has passed, the imagined sound of Nazi jackboots and vision of black-shirted Gestapo functionaries marching up the stairs to his bedroom still occasionally awaken him perspiring from a fitful sleep ([C] p.390). Finally, on July 21, 1938, more than a year after *Das Schwarze Korps* article Himmler sends an official letter to Heisenberg which "cleans him from the accusations". Sommerfeld's chair, however, goes to a mediocre applied physicist: the proponents of "deutsche" physics have obtained in the meantime the support of R. Hess, Hitler's party deputy.

---

[17]Anna Maria Hirsch-Heisenberg has now edited her father's correspondence with his parents: *Werner Heisenberg,, Liebe Eltern! Briefe aus kritischer Zeit 1918 bis 1945*, Langen Müller, München 2003.

[18]It appears surprising that a year after the publication of such an article the British *Nature* would offer its pages to another racist opus of J. Stark (*The pragmatic and the dogmatic spirit in physics*, Nature **141**, April 30, 1938, pp.770-772).

In the summer of 1939 Heisenberg is allowed to go for a month visit to the US. He takes part in a conference in Chicago and lectures at several places on his recent theory of cosmic rays' showers. (His American colleagues are not ready to accept it. According to Heisenberg's recollection of the meeting, the animated discussion following his talk turned into a shouting match between himself and J. Robert Oppenheimer (1904-1967), the doyen of the West Coast physics and the future head of the Manhattan project.) He has turned down the previous year generous offers for professorships at Columbia University in New York and at the University of Chicago. The offers are repeated during his stay and the question is repeatedly raised privately: why does he insist to go back to Hitler's Germany on the eve of a war? He remembers ([H] pp.169-172) a conversation on this topic in Ann Arbor (Michigan). To the arguments of Enrico Fermi (1901-1954) in favour of emigration, Heisenberg responds that he had long since gathered around him a circle of young people whom he saw as a hope for the future, and "if I abandoned them now, I would feel like a traitor. ... I don't think I have much choice in the matter. ... One must be consistent. ... People must learn to prevent catastrophes, not to run away from them. Perhaps we ought even to insist that everyone brave what storms there are in his own country." May be he had later edited his thoughts. Witnesses of this conversation (Goudsmit and Dresden) recall it in a less heroic light – see [C] p.414. In early August, less than a month before the start of World War II, a nearly empty luxury liner (*Europa*) takes Heisenberg back to Germany.

    **5. The Uranium Project**

On September 1, 1939 Hitler's army attacks Poland. Two days later Britain and France declare war on Germany. Werner who was mobilized the year before during the Sudeten crises expects to be called to arms, too. He is indeed mobilized on September 26 – not to the infantry but to the "uranium club" under the auspices of the Army Ordnance. It has been formed to evaluate prospective military applications of uranium fission discovered recently by the Berlin chemists Otto Hahn (1879-1968) and Fritz Strassmann (1902-1980). This is not surprising: physicists on both sides of the coming war have independently alerted their governments to the prospect of a new weapon. (Fermi and Pegram in the US had contacted the US navy, a few months before Heisenberg's visit, about the possibility "that uranium can be used as an explosive". The navy shelved the idea until, later that fall, after the outbreak of the war, Einstein's famous letter, written in August at the urging of Szilard, reached Roosevelt's hands. In Germany the idea was raised in Göttingen and Hamburg and the program was started by Kurt Diebner, at the time director of the Section for Nuclear Physics at the Army Ordnance. Several months later Otto Frisch (1904-1979), a German refugee, and Rudolf Peierls (1907-1995), a student of Heisenberg's, informed the British government on the possibility of a nuclear weapon - [C] p.419.)

Heisenberg starts a theoretical study of the possibilities for exploiting nuclear fission with his usual zeal. Within 3 months he produces the first and in February 1940 the second part of a comprehensive report entitled "The possibility of technical acquisition of energy from uranium fission", establishing himself from the outset as the leading German expert in the field. The eagerness of German physicists to serve their military superiors is not just a manifestation of patriotism (in spite of their disapproval of what they perceived as Nazi excesses) or of the natural scientists' curiosity and their desire to solve a challenging problem. The theoretical atomic physicists, who have for a long time been the butt of deriding accusations that they are busy with vain "Jewish speculations", saw an opportunity to prove themselves useful. That would preserve them from ideological meddling and from reduction of subsidies during the war; it would help to keep gifted young physicists in the lab saving them from the draft to the front. Heisenberg, the Nobel Prize winner, still displays the competitive character of young Werner: once drafted by the Army Ordnance he has to be in charge, not to depend on his unscrupulous (and less intelligent) rivals. In order to be able to support his collaborators and help colleagues in the occupied territories he has to be his own master, and hence, to play to a certain extent the game of those in power. He takes the dubious role of an ambassador of German science to occupied Copenhagen (where the meaning of his meeting with Bohr in 1941 is still debated – see [F] and [PT]) as well as to Krakow, Poland[19], to Holland, to Budapest, and to neutral Switzerland – as a continuing manifestation of the confidence of Nazi authorities towards him. Whatever his intentions, his understanding of the frame of mind of his Western hosts (in Denmark and in Holland) – to whom he spoke of the desirability of a German victory (as opposed to a Soviet occupation) has been totally misguided (see [C] Chapter 22 and [L] p.301 and Secs. 2.4, 2.5).

In 1942 he becomes "director at" the Kaiser Wilhelm Institute in Berlin, a key position, temporarily occupied (after the departure to the US of Peter Debye (1884-1966)) by the party member Diebner (and to which a Lenard-Stark's man from the Ministry of Education has also aspired). For the physicists at the Institute and in the uranium project this appointment is a clear victory of modern over "German" physics. Taking the concurrent professorship in theoretical physics at the University of Berlin requires additional diplomatic efforts by Heisenberg's supporters during what they called "religious debates" (Religionsgespräche) with representatives of the Teacher's League and "deutsche Physik". They reach a 5-point agreement: the party will withdraw from physicists' scientific controversies and Heisenberg will take the professorship, while physicists should refrain from citing Einstein's name in public ([C] pp.454-455). (A quarter of a century ago, Moscow colleagues were making heroic efforts to go around a similar ban on Sakharov's name in the USSR.) Quiet diplomacy and the practical needs of the war have achieved quite a reversal: from the bottom of the "white Jew" affair of 1937, when a theorist could have ended in a concentration camp for merely teaching modern physics, to the appreciation and recognition by the Reich's authorities. The leaders of the NSDAP have changed (somewhat late!) their position toward science giving it priority vis-a-vis ideology. (There is a story, saying that when a session was being prepared in the USSR, soon after World War II, to fight the influence of bourgeois ideology in physics, Igor Kurchatov (1903-1960) had the courage to tell the authorities that they should choose between carrying out such a session and making a (Soviet) atom bomb – and Stalin chose the bomb.)

---

[19]The claim that Heisenberg receives there the Copernicus prize from "his youth friend Hans Frank" - sentenced to death at Nürnberg's trial as a war criminal (made in [C] and recently repeated by J. Bernstein) is been disputed by Helmut Rechenberg (Heisenberg's last doctoral student and curator of the Heisenberg Nachlass at the Max-Planck-Institut für Physik in Munich since 1977). Frank never was a member of Heisenberg's youth movement (according to the available list as well as to Frank's son); Heisenberg was awarded the Copernicus Medal (not prize!) in 1943 by the (old German) University of Königsberg. I thank Dr. Rechenberg for a communications on the subject. See also the September 2004 exchange [Got-B].

Despite claims to the contrary of critics and apologists alike, German physicists are spared facing the difficult moral decision of whether or not to hand a nuclear bomb to Hitler. (This is argued convincingly in [L] on the basis of a critical review of relevant publications like [Wa].) A crucial role is played by the military requirement of December 1941 that a (fully supported) nuclear research effort should produce something of immediate military use within 9 months. After the physicists reply (correctly!) that this would be quite impossible, the uranium project is denied priority and is transferred to the National Research Council. Heisenberg's much discussed June 1942 meeting with Albert Speer, Hitler's newly appointed head of arms production, only confirms the earlier decision to assign a relatively low profile to nuclear research. The German scientists explain that although the construction of a uranium bomb is, in principle, possible, the technical difficulties are so great that its production in Germany in the required period is unrealistic. (A chief difficulty comes from the need for separation of the active isotope U-235 that makes less than 1% of the natural uranium consisting mostly of U-238. Heisenberg argues that a cyclotron should be built in order to do the job. There are two cyclotrons in Germany occupied territory at the time: in Paris and in Copenhagen. It is quite remarkable – as pointed out in the Postscript of [F] – that Heisenberg not only does not try to take over one of them, but actually goes out of his way to protect the one at the Copenhagen Institute in 1944, after Bohr's departure to the US, for the remaining home Danish physicists.) The uranium project should be kept alive anyway since it could lead to the production of energy (through the "uranium machine" – the future reactor). Under these conditions Heisenberg requests – and obtains – for his project a modest funding. (Priority is given to the rockets of Wernher von Braun.) As it has been the case with German mathematicians as well, physicists manage to get a support for their survival from the Nazi government without committing themselves to produce new weapons.

Relegating the uranium project to the back stage, Heisenberg returns to his scientific interests and completes, in the beginning of September 1942 a seminal paper: *On the 'observable quantities' in the theory of elementary particles* ([WH] **AII**,pp. 605-636)*,* in which he develops the theory of a *scattering matrix* (alluded to in earlier work of John Wheeler). Some of the most active theorists studying this concept in the postwar years, like Wentzel and Stückelberg in Switzerland, learned it directly from Heisenberg during his wartime visits – [C] p.478. As with many of his fundamental advances, he has brought to fruition ideas he had discussed (back in the prewar years) with his friend Pauli. It is also in the fall of 1942, in the middle of the war, while relaxing at home, that Heisenberg writes a (161 typescript page long) paper, both personal and philosophical, copies of which he sends at Christmas to a few reliable friends. His words are marked by hard won wisdom and personal courage[20]:

*We have to keep reminding ourselves that it is more important to act humanely than to fulfill any professional, national, or political obligations. ...While in political life there is a constant change of values and the battle of mendacious ideals against other mendacious ideals is unavoidable, in science we enter a realm in which ...there exists a higher power that makes a final decision independent from our wishes...*

He then goes on with words which may sound as an anticlimax but appear important nowadays when governments tend to support almost exclusively applied research: *Most important are, therefore, those areas of pure science that do not lend themselves to talk about practical applications but where pure thought searches for the hidden harmonies of the world.*

Work during the war is hazardous. The Germans are using heavy water – from the Norske Hydro factory in Norway - as a moderator in the 'uranium machine'. In 1943 Norwegian commandos, induced and helped by the British put it out of service. The allied bombings of German towns could no longer be ignored. Heisenberg's weekend's visits from Berlin to his family in Leipzig had to stop after an Allied thousand-bomber raid practically leveled Berlin in early March 1943. The Heisenberg twins, Anna Maria and Wolfgang, also were in Berlin to celebrate Grandfather Schumacher's 75th birthday. Coming from the city centre to the Stegliz suburb with walls of fire on both sides of the smoky road, Werner finds the roof of the Schumacher's house also in flames. Happily his kids and father in law are safe, but he decides to put an end to adventures of this type and moves his wife and children to their Bavarian summer house in Urfeld ([C] p.462). (After another 8 months their Leipzig home is also destroyed.) The time has come to evacuate the Kaiser-Wilhelm Institute as well (although it is still intact – no doubt thanks to St. Florian, the protector from fire, whose statue has been installed by Debye in the corridor – see [EH] Chapter 6 /p.92 of the American edition/). The place for the future reactor is chosen in the rocky caves near Haigerloch, a Western Swabia village.

### 6. The end of the Third Reich. Victors write history

*Luck is a character trait*
Elisabeth Heisenberg

The dangers on Heisenberg's road during the war have been numerous but he seems signed to happy endings.

During his Zürich visit in November 1942 (when he reports on his S-matrix paper) he is followed by agents of both belligerent sides. An agreeable young man is accompanying him to the hotel after a dinner party with colleagues. They continue their animated conversation on the way. Years later, after the war, Heisenberg is presented with a book entitled *Moe Berg, Athlete, Scholar, Spy.* Leafing through it he recognizes in the hero-author the intelligent young Swiss who was sitting in the first row at his lecture and kept a vivid interest in the dinner talk. It turned out that Moe Berg had been a CIA agent, with a loaded pistol in his pocket, and orders to shoot

---

[20] The text, entitled by the editors of Heisenberg's collected work "*Reality And Its Order*" ([WH] **CI**, pp. 217-308) is now made available electronically at http://werner-heisenberg.unh.edu/Ordnung.htm. The author thanks Klaas Landsman for indicating this reference to him.

Heisenberg on the spot at the slightest suspicion that the German physicist was working on the atomic bomb. He writes that, in spite of his special psychological training, he did not notice even a twitch on the suspect face when asking leading questions. A Gestapo agent causes Heisenberg more trouble at the time. He reports that the scientist has made defeatist statements during his visit to Switzerland. Luckily his report is passed to the physicist Walter Gerlach (1889-1979), a party member and deputy of the Reich's Research Council for the uranium project. Gerlach assumes a severe and indignant air and assures the Gestapo man that he will hold Heisenberg responsible. The case ends in a friendly discussion between the two physicists ... ([EH] pp. 97-98).

In December 1942 Heisenberg is invited by the Prussian finance minister Popitz to the Wednesday Society (Mittwochsgesellschaft), an elite Berlin men's club of distinguished intellectuals, academics, statesmen, and military, meeting - since the time of Kaiser Wilhelm – once every several weeks (onWednesdays) in the home of one of its members. Before the refreshments the host used to provide a general lecture on a topic in his field. The 28 members of the Society have been chosen among the German cultural elite that had a critical attitude towards the Nazi regime. By 1944 the idea arose of an assassination attempt on Hitler as an interlude to a military coup. Heisenberg hosts the last meeting of the Society[21] on July 12, at his Institute house in the Berlin suburb of Dahlem (his lecture on the constitution of stars includes an explanation of nuclear fission). Four of the conspirators are among the 10 members attending the meeting. Heisenberg hands the minutes of the meeting and a copy of his lecture to Popitz on July 19 and leaves for Southern Germany stopping for a couple of days at his family house in Urfeld. There they hear the next day, July 20, 1944, on the radio about the failed assassination attempt. The (wounded) war veteran Claus Graf von Stauffenberg had placed a suitcase with a bomb next to Hitler at his headquarters (and gone out). The explosion killed five staff officers but only wounded Hitler. Von Stauffenberg and a few co-conspirators were killed the same night – the first of several thousands, including Popitz and most other members of the Wednesday Society, who have been subsequently arrested, summarily tried (by the "peoples' tribunal"), and executed. Even Planck's younger son Erwin, who has not been present at the July 12 meeting, was not spared[22]. There is no indication that Heisenberg has been even interrogated ([C] p.461). Whether as a result of Himmler's interrogation after Stark's attack in the SS newspaper, or because of his work on the uranium project, his loyalty to the Reich appears then beyond suspicion.

On April 19, 1945 Heisenberg leaves the little village where he had been working with his staff four days before its occupation by the French forces. (A month earlier the physicists have made a last attempt to obtain a chain reaction – without enough safety precaution, risking their lives – but the uranium and the heavy water have not sufficed.) Upon burying the uranium cubes and installing his staff with whatever food supplies remain in the relatively safe basement of a textile factory, he sets out on a bicycle, the only available transport, to his family. After a short stay with his brother he embarks on a reckless bicycle marathon trip across war-torn Southern Germany, a distance of about 250 kilometers. Pedalling at dark only (to avoid marauding German army units and low flying Allied aircraft) he makes it to Urfeld in amazing 3 days. He narrowly escapes death when an SS man looking for deserters stops him the second night and Heisenberg bribes him with a box of American cigarettes (see [EH] p.105 and the dramatic revival in [F] pp.92-93). *And if you do not put your life at stake, your life will be never yours to make*, read his favourite lines of Schiller's *Wallenstein's Camp* (*Und setzet ihr nicht das Leben ein / Nie wird euch das Leben gewonnen sein*).

---

[21] The story is told in complementary details in [EH] and in [C]. The correct date of the last meeting, Wednesday, July 12, is the one in [C]. (The date, July 18, 1944, given in [EH], is a Tuesday.)

[22] This is the last blow to his father, the patriarch of the German physics: he has lost his first wife, the mother of their 4 children (1909), his eldest son, killed in World War I (1916), both his twin daughters dying one after the other while giving birth (1917 and 1919).

On April 30, when the radio announces Hitler's death "on the battle field" (in fact, he is believed to have committed suicide in his Berlin bunker), the happy reunited family opens the last bottle of wine (kept for the baptism of their daughter) to celebrate the event ([EH] p.106). Meanwhile, Colonel Pash from the Alsos mission had arrived (shortly before the French front line) in Haigerloch with a combat engineer battalion. They arrest Heisenberg's men (including von Laue and von Weizsäcker), confiscate their papers, dismantle the pile; by the time the French commanders realize what is happening the uranium and the heavy water are on their way to the Alsos mission headquarters. A few days later heavily armed troops appear on the terrace of the summer house in Urfeld in the pursuit of "target number one": Heisenberg is taken (to an unknown destination) accompanied by tanks and armored vehicles. All Pash is telling his wife is that her husband will be back in 3 weeks. (See the lively description of these events in [EH] p.106.) The prisoners would only return after 8 months (when Heisenberg's mother is no longer alive).

Upon arrival Heisenberg is ushered for interrogation to Goudsmit at Alsos headquarters near Heidelberg. The meeting of the former colleagues is marred by a mutual misunderstanding. For Goudsmit the good feelings and respect to the great physicist have receded and given room to bitterness after Heisenberg's decision to stay in Hitler's Germany and serve its government and his failure to do something tangible when asked to rescue Goudsmit's parents who perished in a Nazi death camp. Heisenberg plays the lucky optimist. Now that the war is over, the horrors of the past should be put behind, old friendships should be restored. He does not show much sensitivity for a man who had been standing a few days earlier in front of the ruins of his house in The Hague and who cannot forget so easily the cruel death of his parents. The impressive operation mounted for his arrest has reinforced his conviction that the American nuclear physicists have much to learn from him. He appears haughty and self-involved. Still Goudsmit greets him with the old question: "Wouldn't you like to come to America now and work with us?" But when Heisenberg repeats his prewar answer, "No, I don't want to leave. Germany needs me.", it sounds as further evidence of his overweening self-importance. (Instead, had Goudsmit tried to understand his interlocutor he might have foreseen the reply: can a person so attached to his country leave it when it is in ruins?) When asked about his work, Heisenberg offers to instruct the Americans on uranium fission. Goudsmit (knowing about Fermi's reactor operating in Chicago since 1942) politely thanks him for the offer. Heisenberg takes the desired as true: he views the interrogation to which he is subjected as a friendly conversation. He writes the same evening to his wife: "The conversations with Goudsmit ... were as amicable as though the last six years had never taken place, and I myself haven't felt so well for years, both emotionally and physically. I am full of hope and ambition for the future. ..." ([EH] p.108; Heisenberg has not been allowed to send letters and his wife finds it after his death among his papers).

There were altogether 10 scientists arrested (including, besides Heisenberg's group, Otto Hahn, Diebner and Gerlach – of those already mentioned). The main reason for their isolation seems to have been the unwillingness of the Americans and the British to share the nuclear secrets with their allies. The knowledgeable German scientists had to be kept out of reach of the French and the Russians. As an American general reportedly expressed the opinion that the best solution to the problem of German nuclear physics was to shoot all German nuclear physicists, the Scottish physicist V.R. Jones, head of intelligence for the British Air Staff, managed to move them to Britain. They were installed at Farm Hall, a sumptuous English country manor in a tiny village near Cambridge (outfitted with secret microphones). When von Laue inquired on the way why they (in particular, he, who had not worked on fission) are held against their will, a British officer replied that they are "*detained for His Majesty's pleasure*" ([C] p.503).

The central event of the Farm Hall transcripts (made available[23] only in 1993) is the detainees' reaction to the broadcast about the Hiroshima atomic bomb on August 6, 1945. They (including Heisenberg) meet the news with disbelief. Later Heisenberg explains his surprise by the fact that Goudsmit had answered his question about the American atom bomb effort by saying, with a smile, that they had more important things to do. "How was I to know that Goudsmit was lying right to my face?" will he complain one day to his wife ([EH] p.109).

The 94 year old Hans Bethe (1967 Nobel Prize in physics, another student of Sommerfeld's who takes part in the Manhattan project in Los Alamos) confirms in [PT] the intuition of the playwright [F]. According to him the Farm Hall transcripts are the best proof, that Heisenberg has not been interested in making a bomb. When the detainees recognize that the American bomb is not a bluff they ask him how could it work? His first attempt at explanation is totally wrong: during the six war years Heisenberg never went to the trouble to calculate the critical mass of U-235, for which an explosion would start. A week later he proves – in another lecture to his fellow detainees - that he has been able to do it: he has worked out an argument, similar to the 1940 theory of (his student!) R. Peierls[24] and O. Frisch, and computes correctly the critical mass. "These two lectures", Bethe goes on ([PT] p.34), "prove that Heisenberg, the scientific leader of the German effort, did not work on a bomb. They show that he did not know critical information and that he could have derived the information if he had tried." Why then did he work on the uranium project? What was he trying to tell Bohr in 1941? Here Bethe offers a personal recollection: "When Bohr came to Los Alamos at the end of 1943, he told Oppenheimer that Heisenberg had talked to him about an atomic bomb. Bohr reproduced from memory a rough drawing that Heisenberg had shown him[25]. The drawing was shown to Edward Teller and me, and we immediately recognized that it is a nuclear reactor with many control rods. ... Perhaps he was trying to get Bohr to be a messenger of conscience, ... to persuade allied scientists also to refrain from working on a bomb." This suggestion (which Heisenberg himself helps diffuse) goes a bit too far. The situation has been better described by a journalist [Po]: "*Zeal was needed,*" Powers says; "*its absence was lethal, like a poison that leaves no trace.*" Heisenberg and his group have worked - selflessly to the last moment, risking at times their lives - with the naive belief that they would impress the victors with a performing reactor – not a bomb.

---

[23] In the introduction to the first publication, [FH], Sir Charles Frank expresses regrets that the transcripts were not released in time for Dürrenmatt to make use of. Bernstein's book [B] has detailed commentaries which are helpful but also prejudiced against the Germans (and against Heisenberg).

[24] Peierls wrote a biographical memoir (with N. Mott) and three essays (all four collected in [Pe]) on Heisenberg and on books concerned with him. Compared with other polemical writings (like [G] [Po] [R] [NYR]) they are concise, skeptical, and lucid.

[25] It is more likely that the drawing was actually shown to Bohr by Jensen who has informed him about Heisenberg's reactor design (K. Gottstein, private communication).

In 1947 Goudsmit publishes his account [G] of the Alsos mission which contains a nasty satire on Heisenberg[26] (as a fierce nationalist, hero worshiped by his men, and having held in contempt more competent reactor physicists like Diebner). American newspapers at the time printed large, boldface headlines: "**TOP-NAZI HEISENBERG**" ([EH] p.111). In letters to Goudsmit (of 1948) Heisenberg only refutes his accusations that German physicists (and he, in particular) had not understood bomb physics. He does not try to justify his moral stand and never cites in his defense his efforts to help Jewish colleagues (we know such cases from people saved by him during the war – see, e.g., Weisskopf's preface to [EH]). As stressed by Landsman [L], Goudsmit is willing to correct in public most of the factual errors of his book, while Heisenberg does not seem to regret anything[27]. He simply does not understand how his courageous behaviour during the Nazi era, contrasted with the comfortable escape of emigration, could possibly be the subject of controversy. In this light Goudsmit's 1976 obituary of Heisenberg is magnanimous: "*Heisenberg was a great physicist, a deep thinker, a fine human being and also a courageous person. He was one of the greatest physicist of our time, but he suffered severely under unwarranted attacks of fanatical colleagues. In my opinion he must be considered to have been in some respects a victim of the Nazi regime.*" ([C] p.522)

Later apologies notwithstanding, the 1947 book [G] continues to serve as a reference point for attacks on Heisenberg[28]. "*History is always written by the victor,*" he once said with resignation to his wife, "*we just have to live with that*" ([EH] p.112).

### 7. Last illusions

After the "fat days of internment" ([EH] p.129) some ten long months have to pass, a time of hardship and depression, before the Heisenberg family is allowed to unite in Göttingen. The British, advised by scientists like Patrick Blackett (1897-1974), former Rutherford assistant and 1948 Nobel Prize winner (who met with Heisenberg during his internment), were willing to help rebuild German science. In a country in debris, torn into four occupation zones with difficult access between them, this is not easy. At times things look hopeless. "I'm totally exhausted" Werner writes to his wife, "it is not just the hunger. I am no longer equal to this continuous organizing with all its disappointments" ([EH] p.130). In spite of everything, he continues to reject offers for work in blissful America, where discouraged colleagues and acquaintances seek to immigrate. He knows well what he is losing: "I am aware that America will be the centre of scientific life during the coming decades, and that the conditions for my work will be much worse in Germany than over there... I, for my part, want to try and help with the reconstruction here ... If the discord of the politicians is not too disturbing, it should be possible to awaken some of the lively intellectual spirit of the 20's" ([EH] pp.128-129). It takes a heroic effort and determination of his wife to arrange by herself for trucks, to gather the furniture from their summer house in the American zone and from Heisenberg's working place (near the cave) in the French, and to move everything – along with the six kids – to the new home at the outskirts of Göttingen given them by the British. There followed months of still hard work to rebuild the Institute, to attract and teach young scientists. But there were also days of active relaxation: playing chamber music with friends, organizing excursions into the countryside with family and colleagues. Shortly before his death Werner will tell his wife that this was the happiest time in his life.

---

[26] Fair, even handed reviews of [G] are contained in [L], pp.304-305, and in the postscript to [F], p.106.

[27] Landsman points out that Heisenberg has missed more than one opportunity of "writing errata to his life"[L], p.322. Heisenberg is also reproved by G. Holton in [PT] for writing in 1955 "that Einstein, to whom war was hateful, should have been moved ... to write a letter to President Roosevelt in 1939, urging that the United States vigorously set about the making of atomic bombs..." which eventually "killed many thousands of women and children". Being more critical to others than to oneself certainly is not peculiar just to Heisenberg. Otto Frisch, one of the participants of the American bomb project, recalls how "Somebody opened the door and shouted, 'Hiroshima has been destroyed!'; about 100000 people were thought to have been killed. I still remember the feeling of unease, indeed nausea, when seeing how many of my friends were rushing to the phone to book tables in the La Fonda Hotel in Santa Fe, in order to celebrate." ([L] pp. 321-322)

[28] A prominent example is provided by the well researched and passionate assault on Heisenberg [R], justly criticized in both [L] and the postscript to [F].

    Heisenberg's activities in the following years are numerous: he is director of a fast growing Max-Planck Institute for Physics and Astrophysics (which moves eventually to Munich in 1958), accepts the position of President of the Göttingen Academy (on its 200 year jubilee); he is particularly interested, however, in creating a German Research council (Deutscher Forschungsrat) or DFR, which is created by the Max-Planck Society in March 1949 with himself as chairman, and in reviving the Alexander von Humboldt Foundation to which he became the first President after the war, in 1953. He feels a responsibility not just of restoring German science but also of creating a healthy political climate in (West) Germany after years of turmoil. DFR has been originally thought of as a panel of 24 selected scientists to provide a critical advice to the government. Scientists and politicians, says Heisenberg, are people with opposite qualities who can fruitfully complement each other. No wonder, politicians are not eager to share their prerogatives with scientists. Chancellor Adenauer is playing for a time the DFR against the rival Notgemeinschaft der Deutschen Wissenschaft (Emergency association of German Scholarship), which represents the interest of Federal states (Länder) keeping the decision making for himself. When the two organizations finally merge (with Gerlach's help) in August 1951 to form the present day German Research Association (DFG), Heisenberg is viewing this as a defeat, albeit he and the DFR managed to get both Federal and some Marshall Plan support as well as the admission of the Federal Republic to UNESCO. As a President of the Humboldt Foundation (a post he cherishes above all the offices he has had - [C] pp.528-529) he implements his ideal that scientists should form an international family with lifelong ties. This is his last official post held for 22 years until illness forces him to resign in 1975 (a year before his death[29]). He helps create the first (and to this day the largest and most important) European scientific laboratory, CERN, in Geneva in 1954, but, true to himself, he refuses the post of Scientific Director there in order to be able to concentrate on his work at home.

    Once giving up the apolitical stand of his youth, in spite of disappointments, Heisenberg does not withdraw from public life. With West German self-rule imminent after 1950, he and fellow nuclear scientists push to establish a cabinet level ministry for nuclear energy policy. At the same time they advise the government to object the NATO plans of equipping the German army with tactical nuclear weapons. When in 1957, in spite of warnings, Adenauer seems to yield to Western pressure, Weizsäcker and (the then ill in bed) Heisenberg compose a Manifesto against the nuclear armament of Germany. Signed by 18 prominent nuclear physicists (including Hahn and Gerlach) the *Göttingen Manifesto* is published (on April 13, 1957) in

---

[29] Two books, [DU] and [PLS], dedicated to Werner Heisenberg, were published in 1977 and in 2002 by the Alexander von Humboldt Foundation. The first opens with the Heisenberg's article *Die Einheit der Natur bei Alexander von Humboldt*, pp.12-23 and contains a paper by the Bulgarian philosopher G. Bratoev, among others.

leading German newspapers and has worldwide reverberations (predating the noisy "anti-nuclear peace movement" – [L] p.300). German scientists strive against nuclear weapons but for the right to exploit nuclear energy, and succeed on both scores. Ensuring that the German army remains non-nuclear, they negotiate with Washington a permission to work on nuclear reactors, a program that by the late 1960's has been one of the most successful in the world ([C] p.512).

By 1955 when the West German Federal Republic is granted sovereignty as part of NATO (and Allied science control laws are rescinded) Heisenberg's main preoccupation becomes once again his work on fundamental problems in quantum physics. The evolution of his views is quite remarkable. At a time when the Landau school in Moscow has pronounced "dead" quantum field theory, and Chew and followers on the West Coast propose, similarly, to use as a fundamental concept, replacing this theory, the *observable scattering matrix* (introduced by Heisenberg during the war), Heisenberg tries to build up a theory of a (not directly observable!) *spinor field*[30], which obeys a non-linear partial differential equation, with a coupling constant of dimension length square. The proton and the neutron, the building blocks of the atomic nucleus, are not identified with the quanta of the basic field (as it is the case with electrons in quantum electrodynamics), but arise as its bound states. Heisenberg does not follow the beaten tracks (not even those that brought him success in the old days), he is looking for something really new. He is experiencing a great uplift. His letter to his wife's sister Edith, written in January 1958, just after having convinced the ever critical Pauli in Zürich that he is on the right track, speaks for itself: "... the last few weeks were full of excitement for me... I have attempted an as yet unknown ascent to the peak of atomic theory, with great efforts during the last five years. And now, with the peak just ahead of me, the whole terrain of interrelationships in atomic theory is suddenly and clearly spread out before my eyes... Not even Plato could have believed them to be so beautiful. For these interrelationships cannot be invented; they have been there since the creation of the world..." ([EH] pp.143-144). Three days before the appearance of the preprint with Pauli (in February 1958) an impatient journalist, present at Heisenebrg's colloquium talk in Göttingen, makes public the word of a new "world formula". Hearing about the preprint, before having received a copy, Landau speaks about Heisenberg's work at his Moscow seminar...

Publicity and commotion are contraindicative to fundamental theoretical advances. Heisenberg (at 57) does not escape the sore fate of Schrödinger (at 60) of 1947, and of Einstein (at 70) of 1949, as their current attempts to construct a unified field theory have been publicized by the press. In fact, Heisenberg is conscientious, that there are questions and "details", which are not yet clear, but he is an optimist. For less emotionally engaged theorists the remaining difficulties and open problems set doubt on the whole enterprise. Shortly before his death later the same year (1958) Pauli withdraws his name from their joint work and expresses publicly his disapproval. Heisenebrg is affected and disappointed but he continues steadfastly his work (with his young collaborator H.P. Dürr). He works, like most great physicists of the past century in their later years, in virtual isolation, with the world of science ignoring his latest ideas.

With the hindsight of subsequent development we can now point to both a prophetic vision and some shortcomings in this last attempt of the great scientist at a new breakthrough in fundamental physics.

---

[30] i.e., a field carrying spin (proper angular momentum) $\hbar/2$; fields with half integer spin are not single valued functions of space time points: they change sign under a rotation of 360 degrees.

The quarks' field in what we now call the standard model can be viewed as a realization of Heisenberg's idea. (He discusses himself such a possibility after the first success of the quark model – some 10 years after his controversial 1958 paper.) In both cases the observable (strongly interacting) hadrons are bound states of the basic field whose quanta do not correspond to free particles. The idea of a spontaneously broken symmetry, associated today with the name of Higgs, plays a prominent part in the Heisenberg model: it originates in his treatment of ferromagnetism of the 1930's. The concept of a fundamental length is present in the now fashionable "superstring theory". (It is traditionally identified there with Planck's length that is many orders of magnitude smaller than the one used by Heisenberg; more recently, however, one also considers larger "compactification radii" in the framework of string theory.) Among the above three cases of late-in-life visions of famous physicists, the one of the 56 years old Heisenberg is probably closest to our theoretical picture half a century later. What is missing in it is the attractive from a geometric point of view *gauge principle*, on which the standard model of strong and electroweak interactions is based, and which secures the renormalizability of the model thus making applicable perturbation theory. Heisenberg is well aware of the non-renormalizability of his model but he hopes that the gauge invariant equations of quantum electrodynamics will follow from his theory at distances large compared to the fundamental length. He does not trust what he views as formal mathematical criteria, but his intuition seems to have failed him at this point.

Deservedly respected in his country, Heisenberg has been an influential figure in postwar Germany. His contribution to restoring the position of science, in general, and of physics, in particular, in the Federal Republic can hardly be exaggerated. We mentioned his role in creating CERN, the only European Laboratory for elementary particle physics, which can compete with those in the US. Yet the next generation of German theorists has its grievances. Heisenberg has been underestimating the new trends in mathematical physics (a tendency reflected in his evening lecture in Trieste [FLP]). His strong personality did not encourage new (independent) developments at his Institute. Heisenberg himself has been disappointed – and often depressed - during the postwar years, both by what he perceived as his failures in science politics and by the reception of his latest work.

Scientists are praised for their top achievements, which are usually (at least for theoretical physicists) the fruit of their youth. (Nobody is praising Einstein for the unsuccessful attempts to create a unified field theory in the last three decades of his life.) The fact that the postwar activity of Heisenberg is seriously scrutinized demonstrates that he has been unusually active until late in life. Observing how theoretical physics is losing positions these days in Germany, we should appreciate the ability of Heisenberg to persuade the authorities of his time to secure the growth of science in a period when the country has been in ruins even more.

Heisenberg was a man with a strong sense of responsibility who lived and worked in difficult times without breaking down. His character and his actions display a unity. Features, which critics, and sometimes even his loved ones, do not like – his obsession with his work and concentration on a single idea, his self-confidence and competitiveness, his strive to be the best at whatever he does - are inseparable from the qualities which had him to his great scientific discoveries. Yet, their relative weight changes with time. Discomforting as it may sound (to us, "the grown-ups"), the 23 years old youngster, who has made a decisive step in the creation of quantum mechanics, hoping for and doubting whether he is on the right track, incites more sympathy than the proud mature man who, 33 years later, is quite confident that the splendid peak of the long sought quantum field theory is shining before his eyes.